\documentclass[english,aps,preprint]{revtex4}
\usepackage[T1]{fontenc}
\usepackage[latin1]{inputenc}
\usepackage{graphicx}
\usepackage{amssymb}

\makeatletter

\providecommand{\tabularnewline}{\\}

\makeatletter

\usepackage{color}

\makeatletter

\newcommand{\bee}{\begin{equation}}
\newcommand{\ee}{\end{equation}}
\newcommand{\beea}{\begin{eqnarray}}
\newcommand{\eea}{\end{eqnarray}}

\makeatother

\makeatother

\usepackage{babel}
\makeatother
\begin{document}

\title{\bf \large \centerline{Localized eigenmodes of the overlap operator and their impact}\centerline{ on the eigenvalue distribution }}

\author{Anna Hasenfratz}

\email{anna@eotvos.colorado.edu}

\affiliation{Department of Physics, University of Colorado, Boulder, CO-80309-390}

\author{Roland Hoffmann}

\email{hoffmann@pizero.colorado.edu}

\affiliation{Department of Physics, University of Colorado, Boulder, CO-80309-390}

\author{Stefan Schaefer}

\email{stefan.schaefer@desy.de}

\affiliation{NIC, DESY, Platanenallee 6, D-15738 Zeuthen, Germany}

\begin{abstract}
In a system where chiral symmetry is spontaneously broken, the low energy
eigenmodes of the continuum Dirac operator are extended.
On the lattice, due to discretization effects, the Dirac operator can have localized
eigenmodes that affect physical quantities sensitive to chiral symmetry.
While the infrared eigenmodes of the Wilson Dirac operator are usually extended even
on coarse lattices, the chiral overlap operator  has many localized
eigenmodes in the physical region, especially in mixed action simulations.
Depending on their density,  these  modes can introduce strong lattice artifacts.
The effect can be controlled by changing the parameters of the overlap operator, in particular 
the clover improvement term
and the center of the overlap projection. 
\end{abstract}
\maketitle

\section{Introduction}

Mixed action simulations with overlap valence quarks on dynamical
configurations generated with a non-chiral fermion action can combine
the advantages of chiral operators in the measurement with relatively
fast configuration generation. The success of this approach depends
to a large extent on how well the valence action matches the sea action.
At the very least the valence action should not introduce any new
large lattice artifacts. In this paper we draw attention to a non-perturbative
lattice artifact, due to localized eigenmodes of the Dirac operator,
that strongly affects the overlap operator, especially in mixed
action simulations.

In the phase where chiral symmetry is spontaneously broken, the low energy eigenmodes
of the continuum Dirac operator are expected to be extended, delocalized.
We have investigated the localization properties of the eigenmodes
of the Wilson Dirac operator and several different overlap operators.
We found that while both the Wilson and overlap operators have localized
eigenmodes, in case of the Wilson Dirac operator these modes usually do not
mix with the infrared modes  but they can become part of the low energy
spectrum of the overlap operator. The source of these modes are dislocations
and the localized modes of the overlap operator can be related to
the localized modes of the kernel operator. The density of these non-physical
overlap modes depend on the kernel operator, on the parameters of
the overlap construction, on the gauge configurations (quenched or
dynamical or mixed action) and on the lattice spacing and can be particularly
large in case of quenched or mixed action simulations.

The physical effect of the localized eigenmodes is easily observable
when the distribution of the Dirac eigenvalue spectrum is compared
to the predictions of Random Matrix Theory (RMT). RMT relies on very
basic assumptions and in quenched systems the only condition for its
validity is that the wave functions of the non-zero Dirac eigenmodes
are extended over the whole volume \cite{Damgaard:2000ah,Verbaarschot:2000dy}.
We show that the  deviations between RMT predictions and the measured
distributions are closely related to the density of
the localized overlap eigenmodes. We argue that some of the
same lattice artifacts are responsible for
the rather large scaling violation effects observed in the topological
susceptibility as well. Our observations suggests that in order to
minimize scaling violations in valence overlap simulations it is not
sufficient to rely on automatic perturbative $O(a)$ improvement but that non-perturbative
lattice artifacts due to dislocations also have to be considered.

\section{Notations and Parameters}
\label{sect2}

We want to study the lattice artifacts due to the localized low energy
eigenmodes of the overlap operator. We consider several different
overlap operators, and in order to facilitate the comparison between
them we calculate their eigenvalues on the same quenched configuration
set, consisting of about 1000 $12^{4}$ configurations with Wilson plaquette
action at $\beta=5.8458$ ($a\approx0.12$fm). 

Our definition of the massless overlap operator is \begin{equation}
D_{{\rm ov}}=R_{0}\left(1+d(d^{\dagger}d)^{-1/2}\right)\;,\quad d=D_{K}-R_{0}\;,\label{eq:Overlap_def}\end{equation}
 where $D_{K}$ is the kernel operator and $R_{0}$ denotes the center
of the overlap projection. We chose $D_{K}$ to be the Wilson operator
with n-HYP smeared gauge connections \cite{Hasenfratz:2001hp,Hasenfratz:2007rf},
both unimproved and with tree level ($c_{\,{\rm SW}}=1$) clover improvement.
These kernel operators were motivated by our ongoing dynamical simulations
\cite{Hasenfratz:2007rf}.

The choice of the parameter $R_{0}$ in the overlap construction is
rather arbitrary, as long as it is larger than the eigenvalues of
the physical, infrared modes of the kernel operator but smaller than
the doubler modes, and the resulting overlap operator is local. Since
the additive mass renormalization and hence the location of the IR
edge of the spectrum varies with the kernel action,
the quantity $\Delta R_{0}=R_{0}-\lambda_{\rm crit}$, with
$\lambda_{\rm crit}$ the location where the IR edge of the complex
spectrum intersects the real axis, characterizes
the overlap operator better than $R_{0}$ itself.
We have chosen two
different $\Delta R_{0}$ values with both of our kernel actions.
The parameters are listed in Table \ref{tab:OV_params}, where for
reference we also give the parameters of the overlap action used in
\cite{Giusti:2003gf} on configurations similar to ours.

\begin{table}
\begin{tabular}{|c|c|c|c|c|c|}
\hline 
Action&
Smearing&
$c_{\,{\rm SW}}$&
$\lambda_{\rm crit}$&
$\,\, R_{0}\,\,$&
$\Delta R_{0}=R_{0}-\lambda_{\rm crit}$\tabularnewline
\hline
\hline 
S1&
n-HYP&
0&
0.30&
1.0&
0.70\tabularnewline
\hline 
S2&
n-HYP&
0&
0.30&
0.7&
0.40\tabularnewline
\hline 
S3&
n-HYP&
1&
0.08&
1.0&
0.92\tabularnewline
\hline 
S4&
n-HYP&
1&
0.08&
0.3&
0.22\tabularnewline
\hline \hline
T1&
thin&
0&
0.90&
1.4&
0.50\tabularnewline
\hline
\end{tabular}

\caption{The parameters of the overlap action considered in this study. $\lambda_{\rm crit}$
is the approximate (real) lower bound of the complex kernel spectrum at $\beta=5.8458$ as explained
in  Sect.\ref{sect2}. \label{tab:OV_params}}
\end{table}

The parameter $R_{0}=1.4$ in \cite{Giusti:2003gf} was chosen by
maximizing the locality of the thin link unimproved overlap action
$T1$. The localization of our overlap actions varies depending on their
parameters, but all of them are local.

\section{The eigenvalue spectrum of the kernel and overlap Dirac operators}

\begin{figure}
\includegraphics[width=9cm]{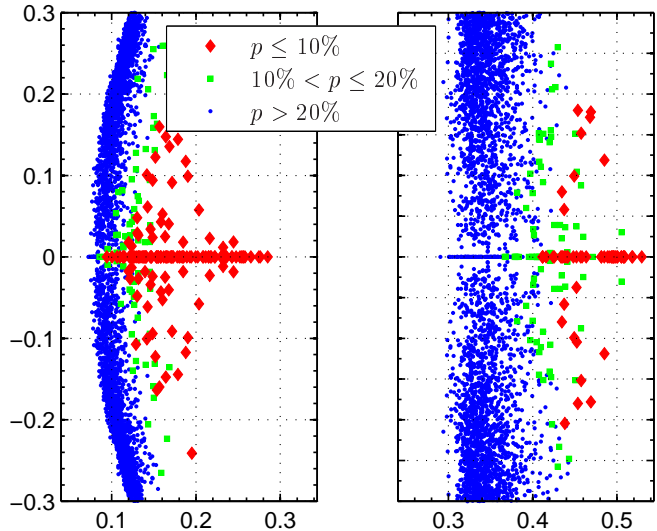}

\caption{The spectrum of the two kernel operators used in this study. Both
are n-HYP smeared Wilson operators, one with tree level $c_{\,{\rm SW}}=1$
clover coefficient (left panel), the other with $c_{\,{\rm SW}}=0$
(right panel). The different plot symbols correspond to different
localization levels of the corresponding eigenvectors . \label{fig:Dirac-spectrum}}
\end{figure}

In this section we study the eigenvalue spectrum and the localization
properties of the eigenmodes of both the kernel and the overlap operators.
Figure \ref{fig:Dirac-spectrum} shows the 40 lowest magnitude eigenvalues
on 100 configurations with both the $c_{\,{\rm SW}}=1$ and $c_{\,{\rm SW}}=0$
kernel actions. (The apparent half-moon shape on the right 
 is only an artifact, due to identifying only the lowest magnitude
eigenvalues.) From Figure \ref{fig:Dirac-spectrum}
we approximate $\lambda_{\rm crit}=0.08$ and 0.30 for the $c_{\,{\rm SW}}=1$
and $c_{\,{\rm SW}}=0$ actions, as listed in Table \ref{tab:OV_params}.
Note that the spectrum of the n-HYP smeared $c_{\,{\rm SW}}=1$ operator
appears much more chiral than the unimproved one, its eigenvalues
are concentrated around a unit circle. This is what makes this action
appealing in dynamical simulations \cite{Hasenfratz:2007rf}. 

A simple and very intuitive measure of the localization of the eigenmodes
is the participation number or inverse of the inverse participation
ratio IPR \cite{Gattringer:2001mn} \begin{eqnarray}
p & = & IPR^{-1}\nonumber \\
IPR & = & V\sum_{x}|\psi(x)|^{4}\,,\label{eq:IPR}\end{eqnarray}
 where $\psi(x)$ is the normalized eigenvector of the Dirac operator.
$p$ varies between $1/V$ and 1, the latter corresponding to a uniform
wave function, the former to one localized on one lattice site. The
participation number is only a qualitative measure: while a very small
$p$ certainly implies a localized mode, a large value does not necessarily
mean a coherent extended one. In addition, the participation number
is sensitive to local fluctuations, so direct comparison is only meaningful
between configurations with similar level of vacuum fluctuations (or
plaquette values). For reference we mention that for a fully
separated smooth instanton--anti-instanton pair of radii $\rho/a=2$
the participation number is $p\approx0.06$ on $12^{4}$ lattices.
Therefore one expects that participation numbers $p<0.1$, probably
even $p<0.2$, correspond to a localized mode.

In Figure \ref{fig:Dirac-spectrum} the different plotting symbols
correspond to different participation numbers of the eigenmodes, and
one observes a strong correlation between $p$ and $\Delta\lambda$,
the radial distance of the eigenvalue from the outer edge of the
approximate circle. Toward
the center of the eigenvalue circle all modes appear to be localized
with small $p$ for both actions. However the spectrum of the clover
improved action has many more localized modes in the vicinity of the
physical, IR range. The spectra in Figure \ref{fig:Dirac-spectrum}
are on $12^{4}$ lattices. We have investigated the localization of
the modes also on $16^{4}$ volumes, though with smaller statistics.
We found that the localization of the kernel operator eigenmodes is
not qualitatively different in larger volumes. The number of eigenmodes
is proportional to the volume, the participation number decreases
with increasing $\Delta\lambda$ and the number of localized modes
in a $1/V$ region around the real axis is approximately independent
of the volume. There are particularly many localized real modes, especially
with $c_{\,{\rm SW}}=1$. 

Since the participation number is strongly correlated with $\Delta\lambda$,
the distance form the infra-red edge of the spectrum, it is meaningful
to define the average participation $\bar{p}(\Delta\lambda)$ as the
average of the participation number of eigenvalues in the vicinity
of the real axis at $\Delta\lambda$. If the typical eigenmodes at
some $\Delta\lambda$ value correspond to extended modes, their average
participation $\bar{p}$ should be volume independent, while in the
region where most of the eigenmodes are localized, $\bar{p}$ will
decrease with the inverse of the volume. Comparing data on $12^{4}$
and $16^{4}$ lattices we see constant $\bar{p}$ values for $\Delta \lambda<0.03$
and $1/V$ dependence for $\Delta\lambda\ge0.05$ for the $c_{\, SW}=1$
spectrum. This finite volume analysis suggests that on the $12^{4}$
lattices eigenmodes with participation number $p<0.40$ at $\Delta\lambda\approx0.05$
are already localized. This in turn indicates that the overlap operator with the
small $\Delta R=0.22$ of action S4 is local since the center of
the overlap projection is still beyond the physical, extended eigenmodes.

The overlap construction {}``projects'' all the modes of the kernel
operator to the Ginsparg-Wilson circle. Of course it does more than
just simply project the modes - opposite chirality modes from the
real axis merge and split into complex modes, and in general the wave
function, and with it the participation number of the modes, can change. 

\begin{figure}
\includegraphics[scale=0.7]{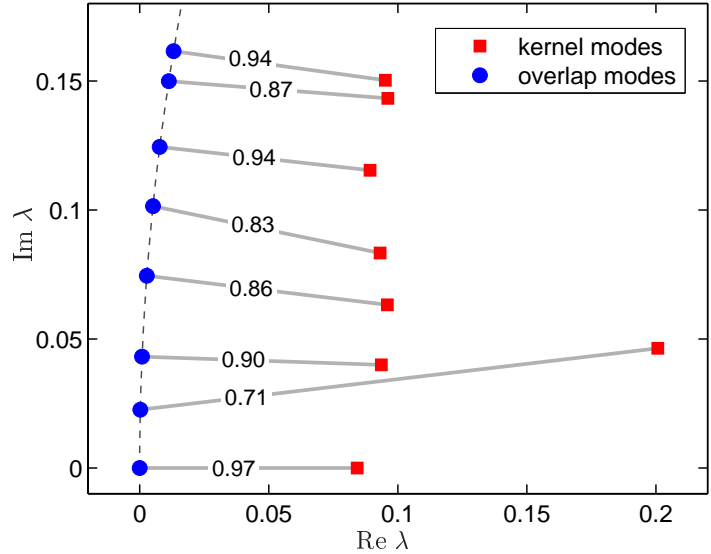}

\caption{Eigenvalues of the kernel and overlap operator on a configuration where
the overlap operator has a localized IR mode. Lines connect overlap modes with
the most similar kernel modes and the magnitude of their inner product is shown.
\label{fig:ov_projection}}
\end{figure}

An example of how the overlap construction transforms the kernel modes
is given in Figure \ref{fig:ov_projection} where a few of the low
energy eigenvalues of the $c_{\, SW}=1$ kernel operator and the corresponding
overlap operator are shown on a configuration with a localized infrared
overlap eigenmode. All but one
of the kernel eigenmodes in Figure \ref{fig:ov_projection} are extended
with large participation numbers. The only exception is the mode in
the inner part of the circle that has participation number $p\approx0.04$.
The eigenmodes of the overlap operator are also extended with one
exception, the mode with the lowest imaginary value has $p\approx0.08$.

The extended overlap eigenmodes all connect strongly to a kernel mode,
with overlap between the wave functions, i.e. the absolute value
of their inner product, of 80\% or larger. The grey
lines in Figure \ref{fig:ov_projection} connect the overlap modes
with the kernel mode with which they have the highest overlap.
It appears that the extended, near infrared eigenmodes change little
under the overlap projection, their eigenvalues basically move out 
straight to the Ginsparg-Wilson circle. This is likely so because  the n-HYP
smeared kernel already has excellent chiral properties and we would
expect the situation to be quite different with an unsmeared kernel.

The localized modes, on the other hand, behave differently.
There is only one localized overlap and one localized kernel mode
in Fig. \ref{fig:ov_projection}. The wave function of both of these
modes are sharply peaked at the same spatial location, they couple
mainly to a small instanton (or dislocation). The overlap of the wave
functions is still sizable, $\approx70$\%, but the eigenvalues are 
different. The overlap eigenvalue is small, it is the most infrared
among the eigenmodes. In general localized modes tend to stay localized
under the overlap projection, their overlap eigenvalue is frequently
small, without modifying the eigenvalues of the extended modes. Hence
these non-physical eigenmodes can strongly influence the low energy
structure of the systems.

\begin{figure}
\includegraphics[height=7cm]{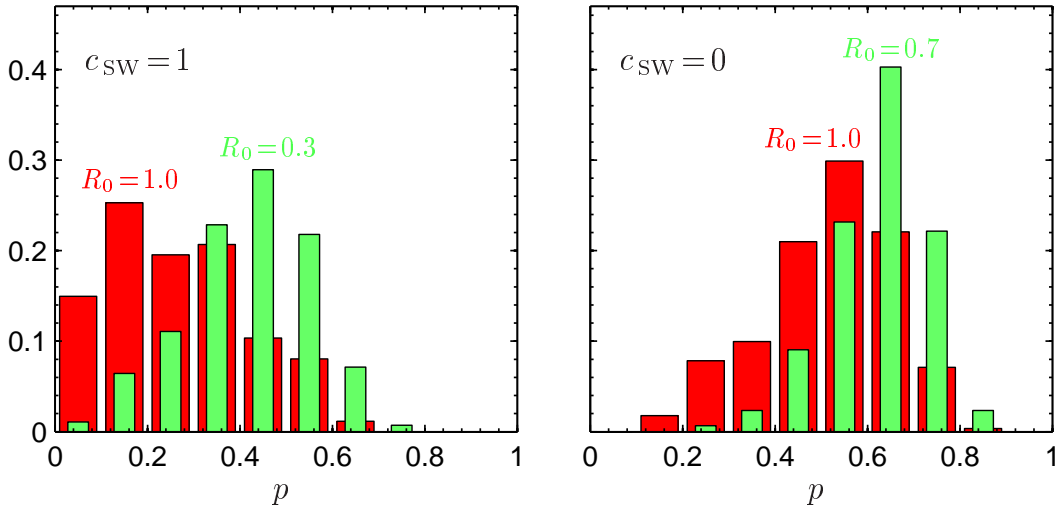}

\caption{The distribution of the participation number of the first non-zero
overlap eigenmodes in the $\nu=1$ sector, normalized by the number
of configurations. \label{fig:IPR}}
\end{figure}

To quantify the observations from Figures \ref{fig:Dirac-spectrum}
and \ref{fig:ov_projection} we have measured the participation number
of the low eigenmodes of our four overlap operators. Figure \ref{fig:IPR}
shows the distribution for the first \textit{non-zero} modes in the
$\nu=1$ topological sector. The result supports what we have expected
based on the eigenmodes of the kernel operator. The S1 action, that
corresponds to $c_{\,{\rm SW}}=1$ improved kernel operator with $R_{0}=1.0$,
has a lot of localized modes - possibly up to 50\% of the first eigenmodes
are localized. The other actions are considerably better. When $R_{0}=0.3$
is used even with the clover improved action, many of the localized
modes are already to the right of the overlap center and projected
to the ultra-violet. Removing the clover term has a similar effect.
Even with $R_{0}=1.0$, corresponding to $\Delta R_{0}=0.7$,
there are only a few localized modes, and their number drops even
further when $R_{0}=0.7$ ($\Delta R_{0}=0.4$) is chosen.

The localized eigenmodes are due to lattice dislocations and their
number scales with the lattice volume. They are non-perturbative
cut-off effects and will make the continuum extrapolation difficult.
We have investigated the localization properties of the overlap eigenvalues
at a finer lattice spacing but same physical volume ($\beta=6.0$
Wilson plaquette action on $16^{4}$ lattices) with the S2 action.
The distribution of the participation numbers of the first non-zero
eigenmodes was slightly worse than the corresponding distribution
on the coarser lattices, making any kind of perturbative predictions
difficult.

We have not investigated the localization of the overlap operator
on larger physical volumes, but based on the spectrum of the kernel
operator, we expect that among the first non-zero eigenmodes of the
overlap operator there are as many localized modes in larger volumes
as in the smaller one. These lattice artifacts are not due to finite
volume effects. The localization of overlap eigenmodes have been studied
recently in Refs. \cite{Ilgenfritz:2007xu},\cite{Hasenfratz:2007yj},
though in different context and only with a single overlap operator.

\section{Consequences of localized overlap eigenmodes}

In the continuum limit the eigenmodes of the Dirac operator are extended.
The localized modes we observed at finite lattice spacing are lattice
artifacts but their presence could influence any physical quantity
that is sensitive to chiral symmetry, like the pion spectrum, the
chiral condensate or the topological susceptibility. Here we study
the latter two quantities.

\subsection{The topological susceptibility}

The topological susceptibility $\chi=\langle \nu^{2}\rangle/V$, when
defined via the index of the overlap operator, is a particularly sensitive
measure of how the fermionic action observes the dislocations of the
vacuum. The index of the overlap is identical to the sum of the chirality
($\pm1$) of the real modes of the kernel operator up to $\lambda<R_{0}$
\cite{Niedermayer:1998bi}. The real modes of the kernel operator
are easiest to identify by measuring the eigenvalues of the Hermitian
operator $\gamma_{5}D_{K}$  and identifying when an eigenmodes crosses zero 
\cite{Edwards:1998wx,DelDebbio:2003rn}.
The advantage of this approach is that one automatically obtains the
topological charge for arbitrary $R_{0}$.

\begin{figure}
\includegraphics[width=10cm]{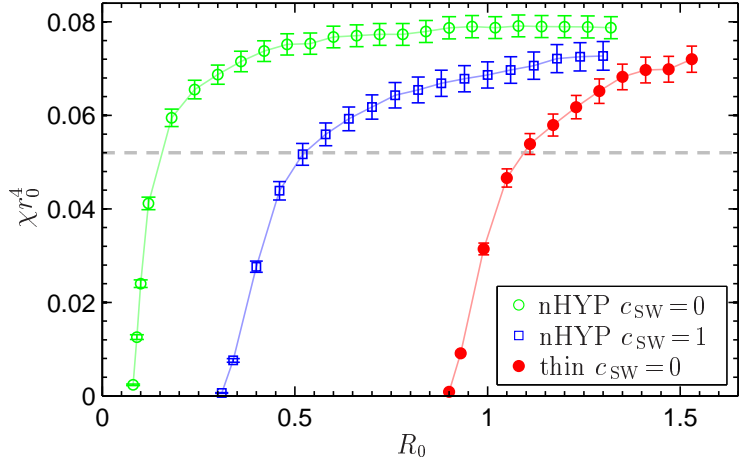}

\caption{The topological susceptibility as a function of the overlap parameter
$R_{0}$ with thin link and n-HYP smeared overlap. The dashed horizontal
line is the continuum prediction from Ref. \cite{Durr:2006ky}. \label{fig:Topo_susc}}
\end{figure}

In Figure \ref{fig:Topo_susc} we show the dimensionless quantity
$\chi r_{0}^{4}$ as a function of $R_{0}$ for the n-HYP smeared
$c_{\,{\rm SW}}=0$ and 1 kernel overlap actions, and also for the
unimproved thin link kernel overlap action. To set the scale we use
$r_{0}/a=4.032$ from Ref. \cite{Necco:2001xg}. We have identified
the real modes of the kernel operators only below $R_{0}<1.4$ (1.6
for the thin link kernel). Had we extended these measurements further,
we would have observed the susceptibility saturate, then drop again
when the right edge of the first eigenvalue circle is approached,
around $R_{0}\approx \lambda_{\rm crit}+2$. Recent studies of the topological
susceptibility using a pure gauge $F\tilde{F}$ topological charge
operator predict $\chi r_{0}^{4}=0.0524(13)$ \cite{Durr:2006ky},
while calculations with a thin link overlap operator give $\chi r_{0}^{4}=0.059(3)$
\cite{DelDebbio:2004ns} in the continuum limit. In Figure \ref{fig:Topo_susc}
we observe not only large cut-off effects, but strong dependence on
the $R_{0}$ parameter, especially for the $c_{\,{\rm SW}}=0$ actions. 

The cut-off effects are the consequence of the large number of real
eigenmodes toward the center of the eigenvalue circle seen in Figure
\ref{fig:Dirac-spectrum}. Most of these modes are lattice artifacts,
dislocations. Overlap operators with smaller $R_{0}$ values are less
sensitive to these inner modes and therefore show smaller lattice
artifacts, but defining the susceptibility where the susceptibility
curve is steeply rising is rather arbitrary. Since $\lambda_{\rm crit}$
depends on the lattice spacing, a small fixed $R_{0}$ value can also
lead to non-monotonic or falsely flat behavior of the susceptibility
as a function of the lattice spacing. Large $R_{0}$, on the other
hand, creates large cut-off effects. Perhaps the most reliable continuum
extrapolation with smallest lattice artifacts would come from where
the parameter $\Delta R_{0}=R_{0}-\lambda_{\rm crit}$ is kept fixed at
a small value  but where the overlap operator is still local.

\subsection{The eigenvalue distribution and the chiral condensate}

The distribution of the low lying Dirac eigenmodes should follow the
universal predictions of Random Matrix Theory for extended eigenmodes.
Localized modes embedded in the IR can spoil the agreement between
the measured and predicted distributions. According to RMT the probability
distribution of the $k$th eigenvalue of the Dirac operator in fixed
topological sector $\nu$ is given as \begin{equation}
p_{\nu,k}(\lambda)=\frac{a}{\Sigma V}\Lambda_{\nu,k}\label{eq:RMT-p}\end{equation}
 with only one free parameter, $\Sigma V/a$, where $\Sigma$ is the
chiral condensate. The universal functions $\Lambda_{\nu,k}$ can
be calculated analytically \cite{Damgaard:2000ah}. In numerical
studies one frequently uses the cumulative or integrated distribution,
\begin{equation}
c_{\nu,k}(\lambda)=\int_{0}^{\lambda}dz\, p_{\nu,k}(z)\,.\label{eq:RMT_cum}\end{equation}

\begin{figure}
\includegraphics[width=14cm]{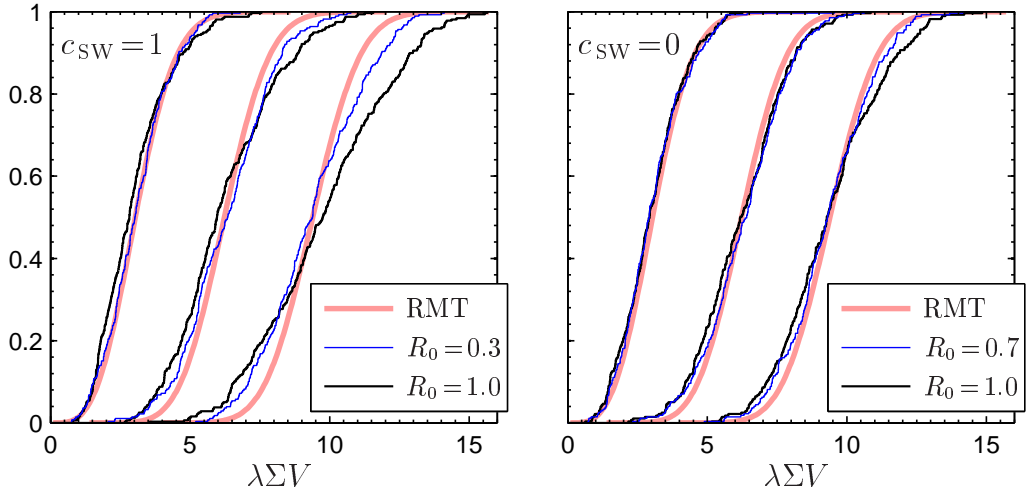}

\caption{The cumulative distribution of the first three eigenmodes in the
$\nu=1$ topological sectors. Left panel: $c_{\,{\rm SW}}=1$; right
panel: $c_{\,{\rm SW}}=0$. The smooth thick lines are the RMT predictions.
\label{fig:cumulative-distribution}}
\end{figure}

Previous studies of quenched QCD found agreement with RMT predictions
when averages $\langle\lambda\rangle$ or ratios of averages were
considered on volumes of at least $(1.5$fm$)^{4}$ for $\nu\le2$
and $k\le4$ \cite{Giusti:2003gf}, but the actual shape of the eigenvalue
distributions shows significant deviations, especially for the higher
modes \cite{Bietenholz:2003mi,Bietenholz:2006fj}.

In Figure \ref{fig:cumulative-distribution} we present our results
for the cumulative distribution using the Dirac operators S1-S4 listed
in Table \ref{tab:OV_params}. We consider only the $\nu=1$ topological
sector, since for all 4 actions close to a third of the configurations
belong there, though the other sectors give similar results. We fix
the free parameter $\Sigma V/a$ by fitting the cumulative distributions
to the RMT predictions. We follow the Kolmogorov-Smirnov procedure
and maximize the quality factor $Q$ ,i.e. the confidence level that
the numerical data is described by the theoretical predictions \cite{RMT}.
The results we present here correspond to a combined fit to the lowest
three modes of a given Dirac operator, but the fit results are very
similar if one includes only one or two of the modes.

In Table \ref{tab:RMT_results}
we give some details of the fit, including the value of $\Sigma V/a$
and the individual quality factors. $Q$ depends exponentially on
the number of configurations and on the square of the maximal deviation
$D_{{\rm max}}$ between the predicted and measured cumulative curves.
When the deviation is systematic rather then statistical, $D_{{\rm max}}$
is actually a better quantity to characterize the fit \cite{Hasenfratz:2006bq}
and we list its value also in Table \ref{tab:RMT_results}. We observe
significant variation in the fitted values of $\Sigma V/a$, even though
the volume is the same for all actions. The renormalized chiral condensate
$Z_{S}\Sigma$ can show cut-off effects that we cannot estimate without
actually calculating the renormalization factors. However a simple
calculation already shows that the $Z_{S}$ factors can be very different
for different overlap operators even when the same smeared kernel
action is used.

To illustrate this point let us model the kernel operator
as $D_{K}=d_{{\rm ov}}+\lambda_{\rm crit}$, where $d_{{\rm ov}}$ is an
overlap operator with eigenvalues $\eta_{\phi}=1-e^{i\phi}$. Since
$d_{{\rm ov}}$ is a normal operator, the eigenvalues $\lambda_{\phi}$
of $D_{{\rm ov}}$ in Eq. \ref{eq:Overlap_def} can be immediately
calculated. The scale factor between $d_{{\rm ov}}$ and $D_{{\rm ov}}$,
$S=\lambda_{\phi}/\eta_{\phi}|_{\phi\approx0}$ depends on the parameters
$R_{0}$ and $\lambda_{\rm crit}$ and is responsible for most of the observed
difference in $\Sigma V/a$. In Table \ref{tab:RMT_results} we list
the values of the scale factor $S$ and $S\,\Sigma V/a$ as well.
It is interesting to note that the ratio $\lambda_{\phi}/\eta_{\phi}$
can show significant dependence on $\phi$ when $\Delta R_{0}=R_{0}-\lambda_{\rm crit}$
is small even for small eigenvalues in the IR range. For us the strongest
dependence is for the action S4 where the effect is $\approx5$\%
for eigenvalues that cover the range we consider here. 
Once the scale factor is taken into account, the predicted values $S\,\Sigma V/a$ 
are consistent for the S1, S2 and S4 actions, only the S3 action with 
the largest deviation $D_{\rm max}$ is statistically different.

\begin{table}
\begin{tabular}{|c|c|c|c|c|c|c|c|c|c|c|}
\hline 
&
&
&
&
&
&
&
&
$n=1$&
$n=2$&
$n=3$
\tabularnewline
action&
smearing&
$c_{\,{\rm SW}}$ &
$R_{0}$ &
$N_{{\rm conf}}$&
$\Sigma V/a$&
$S$&
$S\,\Sigma V/a$&
$\ D_{{\rm max}}\quad Q\quad$&
$\ D_{{\rm max}}\quad Q\quad$&
$\ D_{{\rm max}}\quad Q\quad$
\tabularnewline
\hline
\hline 
S1&
n-HYP &
0 &
1.0 &
279&
$\ $90.8(2.0)&
1.43&
130(3)&
0.056 0.335&
0.081 0.047&
0.090 0.021
\tabularnewline
\hline 
S2&
n-HYP &
0 &
0.7 &
271&
$\ $68.4(1.0)&
1.75&
120(2)&
0.063 0.227 &
0.074 0.095&
0.072 0.118
\tabularnewline
\hline 
S3&
n-HYP &
1 &
1.0 &
318&
143.2(2.4)&
1.09&
156(3)&
0.103 0.002&
0.121 0.000&
0.187 0.000
\tabularnewline
\hline 
S4&
n-HYP &
1 &
0.3 &
279&
$\ $97.9(1.2)&
1.36&
133(2)&
0.039 0.781&
0.082 0.046&
0.098 0.009
\tabularnewline
\hline 
T1&
thin &
0 &
1.4 &
436&
$\ $99.4(2.9)&
2.80&
268(8)&
0.074 0.015&
0.074 0.016&
0.108 0.000
\tabularnewline
\hline
\end{tabular}

\caption{Results of the RMT fit of the cumulative distributions in the $\nu=1$
topological sector. The raw data for the $T1$ action is from Ref.
\cite{Giusti:2003gf}. \label{tab:RMT_results}}
\end{table}

As is evident from Figure \ref{fig:cumulative-distribution} and supported
by the data in Table \ref{tab:RMT_results}, the first eigenmodes
are well described by the RMT curve, but the agreement gets progressively
worse for the higher modes. In general the $c_{\,{\rm SW}}=1.0$ operators
are worse than the unimproved ones. This is surprising as the smeared
kernel action has much better chiral properties and the overlap operator
is also more localized than with the unimproved kernel, however understandable
if the observed lattice artifacts are caused by the localized eigenmodes.

Comparing results of the topological susceptibility (Figure \ref{fig:Topo_susc})
and the eigenvalue distributions we observe that lattice artifacts,
or deviation form the continuum, correlate closely for the two observables
and with the number of localized eigenmodes of the overlap operator
(Figure \ref{fig:IPR}). The action S2 with $c_{\, SW}=0$, $R_{0}=0.7$
deviates the least from the continuum values, the action S3 with $c_{\, SW}=1$,
$R_{0}=1.0$ the most for both quantities. 

\begin{figure}
\includegraphics[width=7cm]{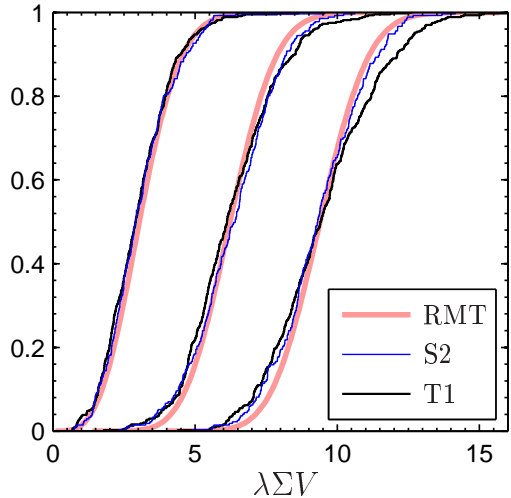}

\caption{Same as Figure \ref{fig:cumulative-distribution} but for the thin
link T1 and n-HYP smeared $c_{\, SW}=0$, $R_{0}=0.7$ S2 actions.
\label{fig:Thin-S2-comp}}
\end{figure}

Figure \ref{fig:Thin-S2-comp} compares the thin link T1 action of
Ref. \cite{Giusti:2003gf} with the S2 data %
\footnote{We thank P. Weisz for sharing with us the data from Ref. \cite{Giusti:2003gf}.%
}. These two actions differ primarily in that S2 is n-HYP smeared.
The smeared data agrees with the theoretical predictions better, though
comparison of the T1 and S1 actions (with larger $\Delta R_{0}$)
shows no significant difference. Further smearing helps very little.
Even with three levels of n-HYP smearing the infrared 
spectrum of the kernel action 
changes little beyond the  reduction of $\lambda_{\rm{crit}}$, and consequently
the cumulative eigenvalue distribution of the S2 action remains 
very similar to the
one with only one level of smearing.

The localized eigenmodes and the deviations they cause are lattice
artifacts, they will disappear in the continuum limit. Nevertheless
at finite lattice spacing they significantly alter the lattice results.

\section{Conclusion and Discussion}

We have investigated the localization properties of different overlap
and their kernel operators in quenched systems. We found that the
overlap operators can have many non-physical, localized eigenstates in the infrared. They
can be related to localized
modes of the kernel operator, but in the kernel operator they are
typically inside the eigenvalue circle and do not directly effect
the low energy spectrum. It is the overlap construction that promotes
them to the infrared.

There their presence can cause
significant scaling violations in quantities sensitive to the properties
of the low modes.
We illustrated that by comparing the eigenvalue distribution of the
low energy eigenmodes to the universal predictions of random matrix
theory and also by investigating the topological susceptibility. One
can minimize the scaling violation effects by choosing a better kernel
operator, like the n-HYP smeared operator we considered here, and
by tuning the $R_{0}$ parameter of the overlap construction as small
as the locality of the overlap operator would allow. Somewhat surprisingly
we found that the kernel operator with the best chiral properties,
the clover improved operator, is actually worse in the overlap construction
as it has the most localized modes near the IR. A chiral kernel operator
reproduces itself in the overlap construction, so it is possible that
other improved kernel operators behave differently. There is indication
that this is indeed the case for the Fixed Point operator in Ref.
\cite{Hasenfratz:2007yj}.

In this paper we considered only quenched systems, but mixed action
simulations suffer from the same problem. The localized modes of the
kernel operator are far from the infrared edge and therefore are not
suppressed by the fermion determinant, yet the overlap can project
them into the infrared. Fully dynamical overlap simulations should
fare better as there the localized eigenmodes are suppressed just
like any other small eigenvalue mode, so while they are present, their
number is at least not inflated. Nevertheless even in dynamical overlap
simulations it is worth minimizing the occurrence of the localized
eigenmodes.

\section{Acknowledgment}

We would like to thank Profs. P. Weisz, P. Hasenfratz, P. Damgaard
and T. Wettig for fruitful discussions. A.H. is grateful for the hospitality
extended to her at the Max Planck Institute in Munich, and for the
opportunity to use the Institute's computer cluster for part of the
calculations presented in this paper. This research was partially
supported by the US Dept. of Energy.

\bibliographystyle{apsrev}
\bibliography{lattice}

\end{document}